\documentclass[useAMS,usenatbib]{mn2e}
\usepackage{graphics}

\newcommand{\be}{\begin{equation}}
\newcommand{\ee}{\end{equation}}
\title[]{Spatial and Velocity clumping in an SDSS blue horizontal branch star catalogue}
\author[Clewley \& Kinman]{L. Clewley \& T. D. Kinman\\
$^{1}$Astrophysics, Department of Physics, Keble Road, Oxford, OX1 3RH, UK; clewley@astro.ox.ac.uk\\
$^{2}$NOAO$^{*}$ P.O. Box 26732, Tucson, AZ 85726; tkinman@noao.edu\\
$^{*}$NOAO is operated by the Association of Universities for Research in Astronomy, Inc.,under cooperative agreement\\
with the National Science Foundation\\} 
\date{Released 2002 Xxxxx XX}

\pagerange{\pageref{firstpage}--\pageref{lastpage}} \pubyear{2002}

\def\LaTeX{L\kern-.36em\raise.3ex\hbox{a}\kern-.15em
    T\kern-.1667em\lower.7ex\hbox{E}\kern-.125emX}

\begin{document}
\label{firstpage}
\maketitle

\begin{abstract}
We present evidence for eight new clumps of blue horizontal branch
(BHB) stars discovered in a catalogue of these stars compiled from the
Sloan Digital Sky Survey (SDSS) by Sirko et al. published in 2004.
Clumps are identified by selecting pairs of stars separated by
distances $\leq$ 2 kpc and with differences in galactocentric radial
velocities $<$ 25 km/s. Each clump contains four or more stars. Four
of the clumps have supporting evidence: two of them also contain
overdensities of RR Lyrae stars which makes their reality very
likely. At least one of the clumps is likely to be associated with the
tidal debris of the Sagittarius dwarf spheroidal galaxy.  We emphasize
that more accurate observations of the radial velocities or proper
motions of the stars in the clumps as well as the identification of
other halo stars in these regions are required to establish the
reality of the remaining clumps.
\end{abstract}


\section{Introduction} 
Hierarchical models of galaxy formation predict that clumps and
streams should be commonplace in galactic halos. Studies of such
structures can give crucial information on the accretion history and
formation of our Galactic halo. Compelling evidence for this came
with the discovery of the disrupted Sagittarius (Sgr) dwarf (dSph)
galaxy (Ibata, Gilmore \& Irwin 1994) and its extensive associated
stream (e.g. Majewski, 2003 and references therein). On a smaller
scale, structure has been found as a ``tidal stream'' associated with
the disintegrating globular cluster Pal. 5. This structure is not
smooth but clumpy (Odenkirchen et al. 2003, Grillmair \& Dionatos
2006).  The origin of this clumping (as a relic of `disk shocking') 
has been discussed by Combes et
al. (1999), Dehnen et al. (2004) and Capuzzo-Dolcetta et al. (2005).

On a still smaller scale, groups of kinematically associated halo
stars have been found by Sommer-Larsen \& Christensen (1987), Doinidis
\& Beers (1989), Arnold \& Gilmore (1992), C\^{o}t\'{e} et al. (1993),
Majewski, Munn \& Hawley (1994), Kinman et al. (1996), Helmi et al.,
(1999), Kundu et al. (2002), Kinman, Saha \& Pier (2004) and Clewley
et al. (2005). Recently, Duffau et al. (2006) have found a group
containing both RR Lyrae and blue horizontal branch (BHB) stars that
they call `The Virgo Stellar Stream' (hereafter VSS). This feature
also coincides with an overdensity of F-type main-sequence stars
(Newberg et al. 2002). Gould (2003) on the other hand, analysed a
large sample of subdwarfs that lie within 300 pc of the Sun and
concluded that not more than 5\% of these stars could be in any one
kinematic group. A good review of the problems involved in detecting tidal
debris in the Galactic halo has been given by Harding et al. (2001).

Blue horizontal branch (BHB) stars are well known tracers of the field
star halo. There is presently only one large well selected survey for
halo objects (Sirko et al. 2004) that we can use to discover halo
structure in both position and velocity space. We therefore use the
1170 BHB star candidates given in Table 3 of Sirko et al. (2004) to
search for these small kinematically-associated groups. We restrict
our search to the 703 stars with $g\leq$18 in order to increase the
probability that the candidates are BHB stars. The distribution of
this sample is shown in Fig. 1.

In section 2 we apply our selection technique to the re-discovery of
the VSS; we conclude that clumps that we find could well be chance
groupings unless there is additional evidence to support their
reality.  In section 3 we use these methods to identify eight new
clumps in the catalogue of SDSS BHB stars given by Sirko et
al. (2004).  We discuss four clumps of BHB stars where there is
additional evidence that they are real and mention four more where
this evidence is lacking. Finally we discuss desiderata for the
discovery of more such clumps and the need for a better understanding
of their association with globular clusters and dSph galaxies.

\section{Techniques for finding streams} 
Our search program takes the galactic coordinates (l, b), the distance
(D) and the galactocentric velocity ($\nu_{los}$) for each star from
Table 3 of Sirko et al. (2004) and calculates the heliocentric X, Y
and Z coordinates for each star \footnote{We use the convention with X
positive towards the Galactic Centre used by Harris, 1996.}. The
distance between each star and the others in the catalogue is
calculated and only those pairs for which this distance is equal or
less than 2 kpc and for which the difference in galactocentric radial
velocity ($\nu_{los}$) is less than 25 km/s are retained.  These
distance and velocity differences were chosen with the aim of
identifying kinematically cold groups of stars out to 30 kpc where the
distance errors are a few percent and the error in $\nu_{los}$ is
about 25 km/s. We describe such sets of stars as being {\it paired};
there are 221 such {\it pairs} of stars in our sample. For the
remainder of this paper we shall refer to this method as the Stellar
Pair Search (SPS) technique.

The VSS contains 8 QUEST RR Lyrae stars (from Vivas et al., 2004) and
5 BHB stars from the SDSS catalogue (Sirko et al., 2004). Its
validity depends upon the coincidence of the galactocentric radial
velocities of its RR Lyrae stars and of the spatial coincidence of
overdensities of both RR Lyrae and BHB stars. Only one of the 221 SPS
{\it pairs} belongs to the VSS and so it is clear that such low
density clumps would not be easily detected in the BHB catalogue by
the SPS method. The principle reason for this is that we are primarily
limited by the accuracy of the velocities in the BHB sample. We are therefore
forced by the accuracy of the velocities to use inter-star distance as
a discriminant which means that we will not find the lower density
streams or structures for which BHB stars are not prominent (like the
VSS or the Pal 5 stream). We therefore concentrate on finding higher density
clumps by considering stars as clump members if they are {\it paired}
with at least {\it two} other members. We further define a clump as
having at least two {\it pairs} of connected stars. There are eight
such clumps in the SDSS catalogue (each having at least four members)
which are listed in Table 1 and we discuss in detail in the following
section. In an attempt to understand the limitations of the SPS
technique we first investigate the probability that these clumps occur
entirely by chance.
\begin{figure}
\centering{
\scalebox{0.45}{
\includegraphics*{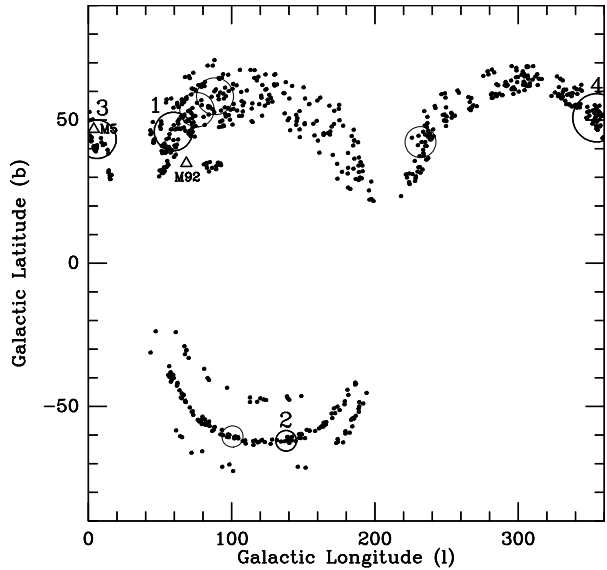}}
}
\caption{{} Galactic coordinates for 703 BHB stars with $g \leq 18$ from Sirko et al. (2004). Circles are
  the location of the eight clumps. Bold numbered circles are the
  locations of the four clumps that have additional evidence that
  supports their reality. The triangles are globular clusters.} 
\label{figure1}
\end{figure}
\begin{table*}   
\begin{flushleft}
\begin{center}
\begin{tabular}{ccrccccccccc}
\hline
\noalign{\smallskip}   
\multicolumn{1}{c}{Clump} &   
\multicolumn{1}{c}{RA} &   
\multicolumn{1}{c}{Dec.} &     
\multicolumn{1}{c}{l} &
\multicolumn{1}{c}{b} &
\multicolumn{1}{c}{$\langle{D}\rangle$} &
\multicolumn{1}{c}{$\langle{R_{gal}}\rangle$} &
\multicolumn{1}{c}{N$_{bhb}$} &   
\multicolumn{1}{c}{$\langle{R}\rangle$} &
\multicolumn{1}{c}{$\rho_{1.5}$} &
\multicolumn{1}{c}{$\langle{\nu_{los}}\rangle$} &      
\multicolumn{1}{c}{L$_{2.0}$} \\
\multicolumn{1}{c}{} &
\multicolumn{1}{c}{h m} &   
\multicolumn{1}{c}{$^\circ$  $^\prime$} &   
\multicolumn{1}{c}{$^\circ$ } &   
\multicolumn{1}{c}{$^\circ$ } &   
\multicolumn{1}{c}{[kpc]} &
\multicolumn{1}{c}{[kpc]} &
\multicolumn{1}{c}{} &   
\multicolumn{1}{c}{[kpc]} &
\multicolumn{1}{c}{[kpc$^{-3}$]} &
\multicolumn{1}{c}{[kms$^{-1}$]} &   
\multicolumn{1}{c}{}    \\
\hline
 1  &    16:14   &    +36:49   &  59.4   & +46.0    &  7.8   & 08.9    &  6  & 0.8  & 0.3 $\pm$ 0.1  & $+$39 $\pm$ 4  &   0.40 \\
 2  &    01:19   &    00:00   & 137.9   & $-$62.0  & 31.6   & 35.2    &  8  & 1.8  & 0.2 $\pm$ 0.0  & $-$36 $\pm$ 6   &   0.82  \\
 3  &    15:33   &    +01:13   & 006.0   & +43.3    & 08.4   & 06.1    &  4  & 0.9  & 0.3 $\pm$ 0.1  & $+$74 $\pm$ 4  &   0.98 \\
 4  &    14:51   &    -00:39   & 354.7   & +50.8    & 09.5   & 07.7    &  6  & 1.3  & 0.3 $\pm$ 0.1  & $-$17 $\pm$ 6    &  0.20  \\
\hline
 5  &    15:27   &    +46:31    & 075.8   & +53.6    & 12.1   & 13.5    &  5  & 1.1  & 0.4 $\pm$ 0.1  & $-$09 $\pm$ 5    &  0.96  \\
 6  &    09:53   &    +05:24    & 231.8   & +42.3    & 09.6   & 15.1    &  4  & 0.8  & 0.3 $\pm$ 0.1  & $+$74 $\pm$ 7   &   0.11 \\
 7  &    23:42   &    $-$00:32    & 088.4   & +58.3    & 07.7   & 11.0    &  4  & 0.8  & 0.3 $\pm$ 0.1  & $-$88 $\pm$ 5  &  0.08 \\
 8  &    00:08   &    +00:04    & 100.5   & $-$60.5  & 15.2   & 17.8    &  4  & 0.9  & 0.3 $\pm$ 0.1  & $-$54 $\pm$ 5   &  0.02\\
\hline
 \end{tabular}
\caption{\label{table1}Clump candidates defined by BHB stars.
              The RA and Dec. are for J2000.
                   $\langle{D}\rangle$ \&
               $\langle{R_{gal}}\rangle$ are the mean heliocentric and
               galactocentric distances respectively. 
               N$_{bhb}$ is the number of BHB stars in each clump and 
               $\langle{R}\rangle$ is their mean distance from the clump 
               centroid. Finally, $\langle{\nu_{los}}\rangle$ is the mean galactocentric
              radial velocity of the clumps. L$_{2.0}$ is defined in the text}   
\end{center}
\end{flushleft}
\end{table*}
\begin{table*}   
\begin{flushleft}
\begin{center}
\begin{tabular}{lccccccccccc}
\hline
\noalign{\smallskip}   
\multicolumn{1}{c}{Sirko no.} &   
\multicolumn{1}{c}{X$_{\odot}$ } &   
\multicolumn{1}{c}{Y$_{\odot}$ } &   
\multicolumn{1}{c}{Z$_{\odot}$ } &   
\multicolumn{1}{c}{R} &   
\multicolumn{1}{c}{V$_{\odot}$} &   
\multicolumn{1}{c}{$\nu_{los}$} &     
\multicolumn{1}{c}{$g_{0}$} & 
\multicolumn{1}{c}{$(g-r)_{0}$} &     
\multicolumn{1}{c}{$(u-g)_{0}$} &
\multicolumn{1}{c}{$\mu_{\alpha}$} &     
\multicolumn{1}{c}{$\mu_{\delta}$} \\
\multicolumn{1}{c}{} &
\multicolumn{1}{c}{[kpc]} &  
\multicolumn{1}{c}{[kpc]} &  
\multicolumn{1}{c}{[kpc]} &
\multicolumn{1}{c}{[kpc]} &
\multicolumn{1}{c}{[kms$^{-1}$]} &   
\multicolumn{1}{c}{[kms$^{-1}$]} &      
\multicolumn{1}{c}{} &      
\multicolumn{1}{c}{} &    
\multicolumn{1}{c}{} &
\multicolumn{1}{c}{[mas/yr]} &    
\multicolumn{1}{c}{[mas/yr]} \\  
\hline  
&&&\multicolumn{3}{l}{}{\bf Clump One}\\
\hline 
   959  &     2.0  &   4.4  &   5.9 &    0.9 &   $-$100  &  +038   &  14.98  &   $-$0.11  &   1.12 & $-$09  & $+$00 \\   
   968  &     2.9  &   4.7  &   5.9 &    0.4 &   $-$102  &  +037   &  15.12  &   $-$0.19  &   1.11 & $-$07  & $-$01 \\   
   972  &     3.2  &   4.8  &   5.5 &    0.5 &   $-$110  &  +034   &  15.05  &   $-$0.21  &   1.16 & $-$08  & $-$01 \\   
   995  &     2.4  &   4.7  &   4.9 &    0.7 &   $-$105  &  +049   &  14.85  &   $-$0.17  &   1.13 & $-$16  & $+$02 \\   
  1110  &     1.8  &   4.5  &   5.7 &    1.0 &   $-$092  &  +051   &  14.95  &   $-$0.16  &   1.11 & $-$04  & $+$01 \\   
  1167  &     4.3  &   4.3  &   5.5 &    1.5 &   $-$106  &  +023   &  15.15  &   $-$0.25  &   1.14 & $-$08  & $-$02 \\   
\hline
  Mean &  2.8 & 4.6 & 5.6 & & &                     +39 $\pm$4 & & & \\
\hline  
&&&\multicolumn{3}{l}{}{\bf Clump Two}\\   
\hline
   548   &   $-$10.2   &   10.4  & $-$28.3   &    0.9  & $-$090   & $-$26  &   18.06  &   $-$0.15  &  1.21 &... &...   \\
   552   &   $-$10.2   &    9.0  & $-$25.4   &    2.8  & $-$125   & $-$65  &   17.85  &   $-$0.15  &  1.18 &... &...   \\
   553   &   $-$10.9   &   10.0  & $-$28.7   &    0.8  & $-$074   & $-$14  &   18.10  &   $-$0.27  &  1.20 &... &...   \\
   556   &   $-$11.1   &   10.4  & $-$29.3   &    1.5  & $-$086   & $-$24  &   18.16  &   $-$0.24  &  1.20 &... &...   \\
   557   &   $-$10.3   &   10.0  & $-$26.8   &    1.2  & $-$110   & $-$46  &   17.97  &   $-$0.10  &  1.20 &... &...   \\
   559   &   $-$12.5   &   10.0  & $-$30.6   &    3.2  & $-$088   & $-$33  &   18.29  &   $-$0.23  &  1.09 &... &...   \\
   562   &   $-$10.7   &    9.4  & $-$25.4   &    2.6  & $-$113   & $-$51  &   17.87  &   $-$0.13  &  1.21 &... &...   \\
   565   &   $-$12.1   &   10.3  & $-$28.9   &    1.6  & $-$094   & $-$33  &   18.14  &   $-$0.17  &  1.20 &... &...   \\
\hline 
  Mean & $-$11.0 & 10.0 &  $-$28  & & &  $-$36 $\pm$ 6  & & & \\
\hline
&&&\multicolumn{3}{l}{}{\bf Clump Three}\\   
\hline
   411   &    5.2  &  0.3  &   6.3  &  1.1   &    063    &  082  &    15.17  &   $-$0.27   &  1.10  & $-$11  & $-$18 \\
   413   &    6.7  &  1.2  &   6.0  &  0.9   &    038    &  079  &    15.35  &   $-$0.20   &  1.18  &  ...  & ... \\
   845   &    6.6  &  0.6  &   5.8  &  0.6   &    037    &  064  &    15.40  &   $-$0.27   &  1.00  & $-$07  & $-$02 \\
   847   &    5.5  &  0.6  &   4.8  &  1.0   &    040    &  070  &    15.00  &   $-$0.23   &  1.00  & $-$16  & $-$12 \\
\hline 
  Mean &   6.0 &  0.7 &  5.7 & & &                       74 $\pm$ 4 & & & \\    
\hline 
&&&\multicolumn{3}{l}{}{\bf Clump Four}\\    
\hline
   337   &   5.7  &  $-$0.5  &  8.0  &  0.7  &  $-$040  & $-$040   &   15.55  &   $-$0.24  &   1.13 & $+$01  & $-$10    \\
   388   &   4.7  &  $-$1.3  &  8.4  &  1.9  &  $-$003  & $-$022   &   15.51  &   $-$0.07  &   1.21 & $-$15  & $-$05    \\
   408   &   5.1  &  $-$0.1  &  6.5  &  1.3  &  $-$020  & $-$010   &   15.31  &   $-$0.18  &   1.16 & $-$06  & $-$16    \\
   812   &   6.0  &  $-$1.3  &  7.4  &  0.8  &  $+$003  & $-$015   &   15.69  &   $-$0.29  &   0.90 & ...  & ...    \\
   833   &   7.4  &  $-$0.2  &  7.1  &  1.5  &  $-$007  & $+$002   &   15.69  &   $-$0.21  &   1.03 & ...  & ...    \\
   837$^{*}$   &   6.6  &  $+$0.3  &  6.4  &  1.4  &  $-$036  & $-$017   &   15.48  &   $-$0.26  &   1.00 & $-$07  & $-$6    \\
\hline 
  Mean  &    5.9 &   $-$0.5 &   7.3 & & & $-$17 $\pm$ 6 & & &   \\
\hline 
\end{tabular}
\caption{\label{table2}Properties of the individual BHB stars in
  clumps one to four summarized in Table 1. X$_{\odot}$,Y$_{\odot}$ \&
  Y$_{\odot}$ and V$_{\odot}$ are the heliocentric coordinates and velocities of the stars. Also listed are the
  de-reddened colours $(g-r)_{0}$, $(u-g)_{0}$ and magnitude $g_{0}$. $^{*}$ The proper motion of this star  is taken from the ACR Catalog around the Celestial Equator (Stone et al., 1999).}
\end{center}
\end{flushleft}
\end{table*}
\begin{table*}   
\begin{flushleft}
\begin{center}
\begin{tabular}{cccccccc}
\hline
\noalign{\smallskip}   
\multicolumn{1}{c}{Clump} &   
\multicolumn{1}{c}{X$_{\odot}$} &   
\multicolumn{1}{c}{Y$_{\odot}$} &   
\multicolumn{1}{c}{Z$_{\odot}$} &   
\multicolumn{1}{c}{$\Lambda_{\odot}$} &
\multicolumn{1}{c}{$B_{\odot}$}   & 
\multicolumn{1}{c}{N$_{1}$}  &  
\multicolumn{1}{c}{N$_{2}$} \\  
\multicolumn{1}{c}{} &   
\multicolumn{1}{c}{[kpc]} &     
\multicolumn{1}{c}{[kpc]} &
\multicolumn{1}{c}{[kpc]} &
\multicolumn{1}{c}{$^\circ$} &
\multicolumn{1}{c}{$^\circ$} &
\multicolumn{1}{c}{}   &
\multicolumn{1}{c}{}   \\
\hline
  1  &$-$04.98 &$-$04.93 &$+$07.37  &224.68 &$-$46.46& 0 &  0 \\
  2  &$-$19.99 &$-$33.12 &$+$05.19  &121.12 &$-$07.65& 0 &  0 \\
  3  &$-$04.02 &$-$03.92 &$+$02.05  &224.24 &$-$20.02& 13&  1 \\
  4  &$-$03.69 &$-$05.84 &$+$01.54  &237.71 &$-$12.57& 16&  3 \\
\hline
  5  &$-$06.01 &$-$08.66 &$+$10.83  &235.21 &$-$45.77& 0 &  0  \\
  6  &$-$15.92 &$-$11.65 &$-$04.80  &216.19 &$-$13.67& 0 &  0  \\ 
  7  &$-$07.94 &$-$07.60 &$+$08.64  &223.75 &$-$38.17& 34&  0  \\ 
  8  &$-$09.51 &$-$17.24 &$+$05.76  &118.89 &$-$16.31& 5 &  2  \\ 
\hline
\end{tabular}
\caption{\label{table3}Relation of clumps to the Sgr stream (Majewski et al.  
 2003). X$_{\odot}$,Y$_{\odot}$ \& Y$_{\odot}$ are the rectangular heliocentric 
 coordinates. $\Lambda_{\odot}$ and $B_{\odot}$ are defined in the text. 
 N$_{1}$ is the number of M-giants (Majewski et al. 2004) that lie within 3.5
 kpc of the centroid of the clump and N$_{2}$ is the number of these whose 
 $\nu_{los}$ lies within 15 km/s of the 
$\langle{\nu_{los}}\rangle$ of the clump.}      
\end{center}
\end{flushleft}
\end{table*}
\subsection{The reliability of the technique}
Harding et al. (2001) suggested that the presence of clumps can be
detected from the lack of normality of the velocity distribution using
the W-statistic of Shapiro \& Wilk (1965). The probability level
(L$_{2.0}$) has been calculated for all the BHB stars within 2.0 kpc
of the centroid of each clump and is given in the last column of Table
1. It should be emphasized that this statistic is quite sensitive to
the size of the sample. Thus, in the case of clump 1, L$_{2.0}$ is
0.40, but if we had taken the stars within 1 kpc of the centroid the
probability level would be 0.08 (a 1 in 12 likelihood that the clump
is a chance grouping).

We investigate the relative probability that such clumps of BHB stars
occur by chance by means of Monte Carlo simulations represented by a
smooth density law falling off as $\rho$ $\sim$ $r^{-3.5}$, where $r$
is galactocentric distance. We assume the simulations mimic the Sirko
et al. (2004) data. Therefore the simulations contain: the same number
of stars, velocity dispersion and volume, i.e. we assume there are 703
BHB stars with g $<$ 18, and a velocity dispersion $\sigma$ = 101
kms$^{-1}$ contained within the Sirko et al. (2004) area.  We also
assume the BHB stars each have exactly the same absolute magnitude. We
recover clumps in the simulations using the SPS method outlined in
\S2. The results of 1000 such Monte Carlo simulations reveal that the
expected number of simulated clumps in one realization is four and
that more than 99\% of the simulations contain seven clumps or
less. The most likely number of stars contained in a clump is also
four although clumps can contain considerably more stars than this;
around 90\% of the clusters contain eight members or less.

It is clear, therefore, that with the survey material at our disposal,
the discovery of clumps in the BHB star distribution can only be a
necessary first step; {\it establishing the reality of these clumps
needs additional evidence}. With these limitations in mind, we now
apply these techniques to the $g \leq$ 18 sample of BHB stars reported
in Sirko et al. (2004) in our search for sub-structure in the halo.
\section{Results}
The SPS technique yielded {\it eight} clumps of BHB stars that contain
{\it four} or more members. Table 1 gives the positions and other data
for these clumps candidates whose space density (denoted as
$\rho_{1.5}$ and defined as the number of stars per cubic kpc within a
radius of 1.5 kpc of the mean clump centroid) is greater or comparable
with that of the VSS.  Additional evidence to support the reality of
clumps 1, 2, 3 \& 4 is given below; data for the individual BHB stars
in these clumps are given in Table 2.

{\it Clump 1}: The centroid of this clump is 1.96 kpc from the
               globular cluster M92. The $\nu_{hel}$ of the cluster
               and clump are +58 and +39 km/s respectively. The clump
               is unlikely to be bound to M92 as there is a
               differential velocity of $\Delta$ V = 19 km/s and it
               would take $\sim$ 10$^{8}$ years for BHB stars to
               travel the $\sim$ 2 kpc from the cluster center. It is
               possible that the BHB stars were tidally stripped from
               the cluster during a recent passage by a stronger
               Galactic gravitational field. Leon, Meylan \& Combes
               (2000) find evidence for such tidal stripping for 20
               globular clusters which can span huge distances
               (e.g. Grillmair \& Johnson 2006). Indeed, there is some
               evidence that M92 is also being tidally stripped of
               stars: Yee et al. (2003) have reported a marginal
               elongation in the NE to SW direction; the clump lies to
               the SW of the cluster.

{\it Clump 2}: This clump of eight stars is contiguous with a clump of
               four BHB stars discovered by Arnold \& Gilmore
               (1992). These authors claim that the stars are not
               bound to each other but may be the stellar remnant of a
               recently disrupted halo cluster.  The centroid of their
               clump is at X$_{\odot}$ = $-$9.1. Y$_{\odot}$ = +11.5
               and Z$_{\odot}$ = $-$29.4 kpc (2.9 kpc from that of our
               clump 2) and has a $\langle{\nu_{los}}\rangle$ of $-$74
               km/s and a standard error of $\sim$5 km/s. Such streams
               of stars can be seen as gradients in the galactocentric
               velocity (e.g. Majewski et al., 2004; Clewley \&
               Jarvis, 2006). Figure 2 plots the V$_{gal}$ versus RA
               for the eight BHB stars in clump 2 and the stars that
               were found to be in a clump by Arnold \& Gilmore
               (1992). The paucity of the data and the large velocity
               errors make this stream detection speculative. Clearly,
               the observation of a larger sample of BHB stars along
               this putative stream is required to confirm its reality.
    
{\it Clump 3}: The centroid of this clump is 1.0 kpc (17 tidal radii)
              from the Oosterhoff type 1 globular cluster M5; its
              $\langle{\nu_{los}}\rangle$ of $+$74 km/s is the same
              as that of the cluster (+75 km/s). Eight RR Lyrae
              stars (QUEST 392, 408, 418, 423, 426, 429, 436, and 456
              (Vivas et al. 2004) lie within 1.5 kpc of the centroid
              of this clump. The period distribution of the type {\it
              ab} RR Lyraes in this group is consistent with their
              belonging to Oosterhoff type 1.  

{\it Clump 4}: Five RR Lyrae stars (QUEST 331, 341, 365, 376 and 386
            lie within 1.5 kpc of the centroid of this clump. Three M
            giants (1411221-061013, 1428255-082436 and
            1434332-0140570) with $\nu_{los}$ of $-$16.8, $-$40.5 and
            $-$7 km/s respectively (Majewski et al., 2004) lie
            within 3.5 kpc of the centroid of this clump. An
            association with the Sgr tidal stream (described below)
            seems likely.

\begin{figure}
\centering{
\scalebox{0.35}{
\includegraphics*{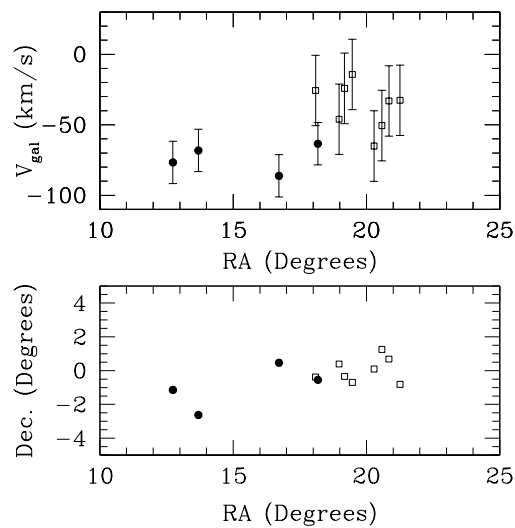}}
}
\caption{{\em Upper:} Plot of V$_{gal}$ versus RA for the eight BHB
  stars in clump 2 (open squares) and four BHB stars found to be in a
  clump by Arnold \& Gilmore (1992). {\em Lower:} A plot of RA versus
  Dec. for the BHB stars.}  
\label{figure2}
\end{figure}
Of the four clumps that have additional observational evidence we
consider clump 4 to be the most convincing with clumps 1,2 and 3
requiring either additional or more accurate observational data. In
an attempt to find such evidence we also investigate the use of available
proper motion data for the clumps. Such proper motions - taken
predominantly from the Second US Naval Observatory CCD Astrograph
Catalog (UCAC2) catalogue - are given in Table 2 and are available for
the BHB stars in clump 1, 3 and 4. In each clump the dispersion in
these proper motions is no more than that to be expected from their
likely errors but this only weakly constrains the possible tangential
velocities of these stars. In short, the UCAC2 proper motion errors
are too large for them to be presently useful in this study.

\subsection{The Sgr stream}
Many types of halo field stars have been found to be associated with this 
stream: M-giants (Majewski et al. 2003, 2004; Law et al. 2005),
RR Lyrae stars (Ivezi\'{c} et al. 2000; Vivas et al., 2001; Vivas et al.,
2005), BHB stars (Newberg et al., 2002; Clewley et al., 2006),
metal-poor K giants (Kundu et al. 2002) and carbon stars (Totten \& Irwin, 
1998). It is therefore desirable to see if any new clumps are associated with
this stream. Majewski et al. (2003) have defined a coordinate system in 
which the zero plane of the latitude coordinate $B_{\odot}$ coincides with the
best-fit great circle defined by a sample of 1161 M-giants, as seen from the
Sun; the longitudinal coordinate $\Lambda_{\odot}$ is zero in the direction of   
the Sgr core and increases along the Sgr trailing stream. About 60\% of these
M-giants lie within 2$\sigma$ (3.6 kpc) of the plane. The coordinates of our
clumps in this system are given in Table 3; only clump 4 seems likely (and clump
3 possibly) to be associated with the Sgr stream. 

\section{Discussion}
The angular diameters of these clumps are comparable with the angular
extent of the survey material in which they were found (Fig. 1).
Consequently they (and the VSS) must extend over distances of at least
a few kpc or more. By comparison, the width of the tidal tails of the
globular cluster Pal. 5 studied by Odenkirchen et al. (2003) is
only about 200 pc. Vivas et al. (2003) found an overdensity of RR
Lyrae stars near Pal. 5 that ``extend even farther away from the
cluster and over a wider range of directions than the narrow tidal
tails''. However, Grillmair and Dionatos (2006) have noted that BHB
stars do not occur in the directions of this cluster's tidal tail and
we can find no compelling evidence for the association of any of our
sample of BHB stars with this cluster. The large extension of these
clumps is understandable if they are the tidal debris of a dSph
galaxy; less so if they come from a globular cluster (as in the case
of clumps 1 and 3).  Presumably there are red giants associated with
clumps 1,3 and 4 that are bright enough for their [$\alpha$]/[Fe/H]
ratio (an indicator of their progenitor's origins) to be obtainable.

In summary, we have used a simple selection technique to isolate eight
clumps of BHB stars.  There is some additional evidence to support the
reality of four of these clumps.  In the case of two of them, the
additional presence of an overdensity of RR Lyrae stars makes their
reality likely.  Further support for the reality of these clumps
could come from improved radial velocities since their expected
velocity dispersions should be small. Further observations of the
radial velocities or proper motions of the stars in the clumps as well
as isolating other halo stars in these regions are required to
establish the reality and nature of the remaining clumps.

\section*{Acknowledgements} 
LC acknowledges PPARC for financial support. This research has made
use of the SIMBAD database, operated at CDS, Strasbourg, France. We
thank Caroline van Breukelen for the use of her various IDL routines. We also
thank the anonymous referee for helpful suggestions to the manuscript.


\begin{thebibliography}{}
\bibitem{}   Arnold, R., Gilmore, G. 1992, MNRAS, 257
\bibitem{}   C\^{o}t\'{e}, P., Welch, D., Fischer, P., Irwin, M. 1993, ApJ, 406, L59,
\bibitem{}   Capuzzo-Dolcetta, R., Di Matteo, P., Miocchi, P. 2005, AJ, 129, 2005
\bibitem{}   Combes, F., Leon, S., Meylan, G. 1999, A\&A, 352, 149
\bibitem{} Clewley, L., Warren, S. J., Hewett, P. C., Norris, J. E., Wilkinson, M. I., Evans, N. W., 2005, MNRAS, 362, 349
\bibitem{}   Clewley, L., Jarvis, M., 2006, MNRAS, 368, 310
\bibitem{}   Dehnen, W., Odenkirchen, M., Grebel, E.K., Rix, H.-W. 2004, AJ, 127, 2753
\bibitem{}   Doinidis, S., Beers, T. 1989, ApJ, 340, L57
\bibitem{}   Duffau, S.,Zinn, R., Vivas, A.K., Carraro, G.  et al., 2006, ApJL, 636, L97
\bibitem{}   Gould, A. 2003, ApJ, 592, 63
\bibitem{}   Grillmair, C.J., Johnson R., 2006, ApJ, 639, 17
\bibitem{}   Grillmair, C.J., Dionatos, O., 2006, ApJ, 641, 37
\bibitem{}   Harding, P., Morrison, H., Olzewski, E., Arabadjis, J. et al. 
       2001, AJ, 122, 1397  
\bibitem{}   Harris, W.E. 1996, AJ, 112, 1487
\bibitem{}   Helmi, A., White, S., de Zeeuw, P., Zhao, H. 1999, Nature, 402, 53
\bibitem{}   Ibata, R.A., Gilmore, G., Irwin, M.J. 1994, Nature, 370, 1941
\bibitem{}   Ivezi\'{c}, Z., Goldston, J., Finlator, K., Knapp, G. et al.  2000, AJ, 
       120, 963 
\bibitem{}   Kinman, T.D., Pier, J.R., Suntzeff, N.B.,Harmer, D.L. et al., 1996, AJ, 111,1164
\bibitem{}   Kinman, T.D., Saha, A., Pier, J.R., 2004, ApJ, 605, L25
\bibitem{}   Kundu, A., Majewski, S.R., Rhee, J., Rocha-Pinto, H.J. et al. 2002, ApJ, 576, L125
\bibitem{}    Law D. R., Johnston K. V., Majewski S. R., 2005, ApJ, 619, 807
\bibitem{}   Majewski, S.R., Munn, J.A., Hawley, S.L., 1994, ApJ, 427, L37
\bibitem{}   Majewski, S.R., Skrutskie, M., Weinberg,M., Ostheimer, J. 2003, ApJ, 599, 1082
\bibitem{}   Majewski, S.R., Kunkel, W., Law, D., Patterson, R. et al., 2004, AJ, 128, 245
\bibitem{}   Newberg, H., Yanny, B., Rockosi, C., Grebel, E. et al. 2002, ApJ, 569, 245
\bibitem{}   Odenkirchen, M., Grebel, E.K., Dehnen, W., Rix, H.-W. et al. 2003, AJ, 126, 2385
\bibitem{}   Shapiro, S.S., Wilk, M.B.  1965, Biometrika, 52, 591 
\bibitem{}   Sirko, E., Goodman, J, Knapp, G.R., Brinkmann, J. et al., 2004, AJ, 127, 899
\bibitem{}   Sommer-Larsen, J., Christensen, P. 1987, MNRAS, 225, 499
\bibitem{}   Stone et al., 1999, AJ, 118, 2488
\bibitem{}   Totten, E., Irwin, M. 1998, MNRAS, 294, 1
\bibitem{}   Vivas, A.K., Zinn, R., Andrews, P. Bailyn C. et al. 2001, ApJ, 554, 33
\bibitem{}   Vivas, A.K., Zinn, R. 2003, Mem. Soc. Astron. Italiana, 74, 928
\bibitem{}   Vivas, A.K., Zinn, R., Abad, C., Andrews, P. et al. 2004, AJ, 127, 1158
\bibitem{}   Vivas, A.K., Zinn, R., Gallart, C. 2005, AJ, 129, 189  
\bibitem{}   Yee, H. K. C., Ellingson E., 2003, ApJ, 585, 215
\end{thebibliography}
\end{document}